\documentclass[a4paper,twocolumn,11pt,accepted=2017-05-09]{quantumarticle}
\pdfoutput=1
\usepackage[utf8]{inputenc}
\usepackage[english]{babel}
\usepackage[T1]{fontenc}
\usepackage{amsmath}
\usepackage{hyperref}
\usepackage{tikz}
\usepackage{lipsum}
\usepackage[numbers,sort&compress]{natbib}
\setcitestyle{numbers,sort&compress} 

\begin{document}

\title{Nonreciprocal photon transport in indirectly coupled whispering-gallery mode resonators}

\author{Gang Li}
\author{Ying Qiao Zhang*}
\email{yqzhang@ybu.edu.cn}
\author{Xing Ri Jin*}
\email{xrjin@ybu.edu.cn}
\affiliation{Department of Physics, College of Science, Yanbian University, Yanji, Jilin 133002, China }

\maketitle

\begin{abstract}
We explore the reflection and transmission characteristics of a system composed of two whispering-gallery mode resonators. Each resonator contains a Zeeman split quantum dot and side-couples to an optical fiber. The results indicate that the unidirectional reflectionlessness and unidirectional transmissionlessness can be achieved by adjusting the coupling strength between the resonators and optical fiber.
Besides, a one-to-one correspondence is set up between the resonance frequencies of quantum dots' energy levels and the positions of low reflection (transmission) peaks when phase shift is $\pi$. This research provides the valuable insights for the developments of quantum optical devices such as isolators, circulators, and routers.
\end{abstract}

Nonreciprocal photonic devices\cite{Guo2009,Calabrese2020,Cheng2012}, such as isolators\cite{WOS:000322450200002}, filters\cite{Ashley2021} and like-diodes\cite{Gu2016}, are commonly used to control photon transport and play the essential roles in quantum information and quantum computing\cite{Lin2010,Chen2009,Wei2011}. However, conventional magneto-optic systems struggle to produce nonreciprocal devices that meet the demands of modern miniaturization. To solve this problem, some alternative non-magnetic techniques are proposed, including dynamic
modulation dependent on refractive index\cite{Williamson2020}, stimulated polarization inter scattering based on photoacoustic effects\cite{Bashan2021}, modulation dielectric constant\cite{Shaikh2016,Grote2001} and  intensity dependent isolation caused by optical nonlinearity\cite{Zhou2017,Hamann2018}. Not only that, several non-magnetic systems such as
metamaterials\cite{Kang2014,Alaeian2014,Yin2023}, optomechanical circuits\cite{Shang2019,Xu2016,Yan2019}, cavity-based photonic devices\cite{Wiersig2014,Zhang2015,Lee2009} and whispering-gallery mode (WGM) systems\cite{Ward2011} have been proposed for the study of nonreciprocal photon transmission.
Among these systems, WGM resonator has garnered a great deal of attention from researchers and been widely employed to investigate photon transport behaviors such as nonreciprocity\cite{Calabrese2020,Lenferink2014}, nonlinear optical response\cite{Chen2019,Yang2017} , electromagnetic-like induced
transparency\cite{Strekalov,Wang2016,Huang2022}, and so on, due to its simplicity in fabrication, low mode volume, high quality factor and minimal energy loss.

In recent years, lots of researches based on WGM resonator focused on exploring the transmission characteristics in nonreciprocal systems\cite{Naweed2005,Mi2011,Chang2014,Jiang2016,Ruesink2018,Ruesink2016}. For example, in 2011, Mi $et~al.$ successfully achieved asymmetric transmission and reflection by using a WGM resonator system embedded with Zeeman split quantum dot\cite{Mi2011}. They also analyzed the effects of some system parameters on scattering characteristics. In 2014, Chang $et~al.$ designed a parity-time symmetric WGM system with balanced gain and loss, which 
enabled the experimental realization of nonreciprocal light transmission\cite{Chang2014}.
In 2016, Jiang $et~al.$ proposed a single WGM resonator with gain-induced nonlinearity to effectively suppress signal power loss, resulting in successful experimental realization of nonreciprocal light transmission\cite{Jiang2016}. In 2018, Ruesink $et~al.$ demonstrated the nonreciprocal circulation of light through radiation pressure interactions in a three-mode system\cite{Ruesink2018}.
The aforementioned investigations mainly focused on transmission characteristics, whereas optical devices based on the mechanism of unidirectional reflectionlessness (UR)\cite{Yang2020,Yang2021,Yu2020} can not only protect upper stage optical devices from reflected light in the cascade circuit to improve system stability, but also convey information, similar to transmission. Up to now, UR has demonstrated the significant potential applications in sensors, diodes, isolators, filters, and so on, and thus researches on UR continues to grow.
In 2021, Huang $et~al.$ employed a pair of indirectly coupled WGM resonators to achieve multi-band UR of photons through modulation of the intermode backscatterings of resonators\cite{Huang2021}. In the reciprocal system that the equivalent transmission in both directions was exhibited. Obviously, these systems previously mentioned only investigated the transmission or reflection characteristics, without considering achieving complete nonreciprocity in both channels, whereas directional transport in both channels is vital to enhance the controllability of photons.

To this end, we propose a non-reciprocal system consisting of two WGM resonators that are individually embedded with a Zeeman split quantum dot (QD)\cite{Zhao2001,Lee2010,Xiao2021} and indirectly coupled through an optical fiber.
By optimizing some system parameters, we demonstrate the simultaneous realization of UR and unidirectional transmissionlessness (UT). Moreover, the conversion between UR and UT can be achieved by adjusting the coupling strength between WGM resonators and optical fiber. Additionally, a one-to-one correspondence is established between the resonant frequencies of QDs energy levels and the positions of UR and UT peaks.

\begin{figure}[t]
\includegraphics[width=3.2in]{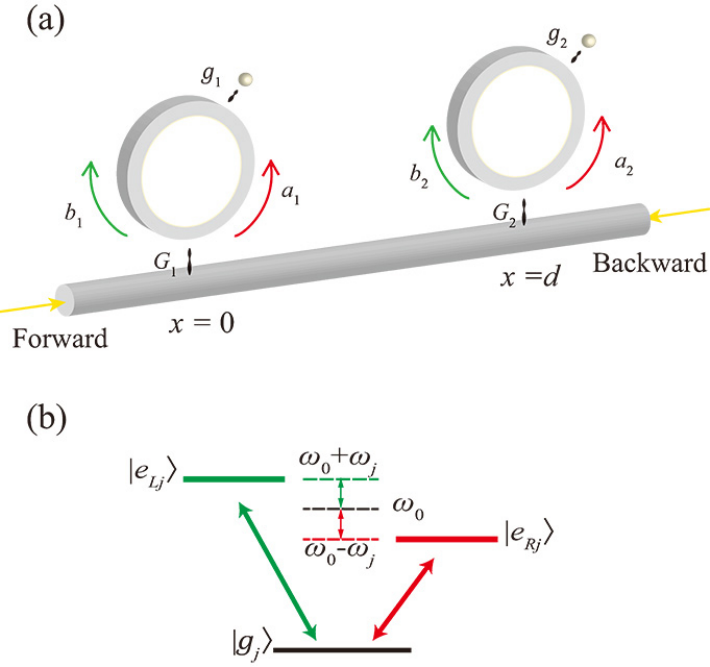}
\caption{\label{1}~(a) Schematic of system under consideration. Two WGM resonators are positioned at $x=0$ and $x=d$ along the fiber, respectively, and are coupled to the fiber with coupling strength  $G_j$ (subscript $j=1,2$~means the first and second WGM resonator). Each resonator inherently supports both clockwise (CW) mode $b_j$ and counterclockwise (CCW) mode $a_j$ which can be converted to each other with a transition rate $h_j$. A Zeeman split QD is coupled to a WGM resonator with coupling strength $g_j$. (b) Schematic of energy levels of each QD. $|g_j\rangle$ is ground state, $|e_{Lj}\rangle$~and~$|e_{Rj}\rangle$~are left- and right-polarized excitons states of the $j$th QD with resonance frequencies $\omega_0+\omega_j$ and $\omega_0-\omega_j$, respectively, where $\omega_0$ is reference frequency. The CW mode $b_j$ and CCW mode $a_j$ only couple to the left-polarized exciton $|e_{Lj}\rangle$ and right-polarized exciton $|e_{Rj}\rangle$, respectively.
}
\end{figure}
\section{Model and calculations}\label{sec:1}
The schematic of system is shown in Fig.\ref{1} (a) and energy levels of QD is shown in Fig.\ref{1} (b). Assuming that the WGM resonators and QDs have the same loss rates $\gamma$, then the Hamiltonian of the system can be written as (assuming~$\hbar=1$)
\begin{widetext}
\begin{align}
     H=&\int dx\{[-iv_gC^{\dag}_{R}(x)\frac{\partial}{\partial x}C_{R}(x)+iv_{g}C^{\dag}_{L}(x)\frac{\partial}{\partial x} C_{L}(x)]\nonumber
     \end{align}
     \begin{align}
         &+{\sum_{j=1,2}G_j\delta[x-(j-1)d](C^{\dag}_RC_{a_j}+C^{\dag}_LC_{b_j}+{\rm H.c.})}\}\nonumber
     \end{align}
     \begin{align}
     &+\sum_{j=1,2}\{[g_j(C_{a_j}\sigma^{\dag}_{Rj}+C_{b_j}\sigma^{\dag}_{Lj})
     +h_jC^{\dag}_{a_j} C_{b_j}+{\rm H.c.}]\nonumber
     \end{align}
     \begin{align}
         &+(\omega_{0}-\omega_{j}-i\gamma)\sigma^{\dag}_{Rj}\sigma_{Rj}+(\omega_{0}+\omega_{j}-i\gamma)\sigma^{\dag}_{Lj}\sigma_{Lj}\nonumber
     \end{align}
     \begin{align}
      &+(\omega_{a_j}-i\gamma)C^{\dag}_{a_j}C_{a_j}+(\omega_{b_j}-i\gamma)C^{\dag}_{b_j}C_{b_j}\},
\end{align}   
\end{widetext}
where~$C^{\dag}_{R}(x)~(C_{R}(x))$~and~$C^{\dag}_{L}(x)~(C_L(x))$~are creation (annihilation) operators at $x$ for forward and backward propagating photon along fiber, respectively.
$C^{\dag}_{b_j}~(C_{b_j})$ and~$C^{\dag}_{a_j}~(C_{a_j})$~are creation (annihilation) operators of CW mode $b_j$ and CCW mode $a_j$ with resonance frequencies $\omega_{bj}$ and $\omega_{aj}$, respectively.
$\sigma^{\dag}_{Lj}~(\sigma_{Lj})$ and $\sigma^{\dag}_{Rj}~(\sigma_{Rj})$ are
transition operators which mean the transitions from states $|g_j\rangle~(|e_{Lj}\rangle)$ to $|e_{Lj}\rangle~(|g_j\rangle)$ and $|g_j\rangle~(|e_{Rj}\rangle)$ to $|e_{Rj}\rangle~(|g_j\rangle)$, respectively.
$v_{g}$~is group velocity of photon.
$G_j~(g_j)$ is coupling strength between the $j$th WGM resonator and fiber ($j$th QD), and $h_j$ is transition rate between $b_j$ and
$a_j$. We set up $\omega_{a1}=\omega_{a2}=\omega_a$, $\omega_{b1}=\omega_{b2}=\omega_b$, $G_1=G_2=G$, $g_1=g_2=g$~and $h_1=h_2=h$ for the sake of simplicity in the following discussions.

Assuming that photon with energy $E_k=\omega=v_g k $ is incident along the forward direction, where $\omega$ and $k$ are frequency and wave vector of the incident photon, and state of system is (assume $\omega_a=\omega_b=\omega$ and system is originally prepared in vacuum state~$|{\rm O}\rangle$)
\begin{align}\label{e002}
|\Psi\rangle
 =& \int dx[\Phi_{k,R}(x)C^{\dag}_{R}(x)+\Phi_{k,L}(x)C^{\dag}_{L}(x)]
|{\rm O}\rangle\nonumber \\
&+(\varepsilon_{aj}C^{\dag}_{a_j}+\varepsilon_{bj}C^{\dag}_{b_j}+\xi_{lj}\sigma^{\dag}_{lj})|{\rm O}\rangle,
\end{align}
where~$\Phi_{k,R}(x)$~and~$\Phi_{k,L}(x)$~represent the wave functions at $x$ of the fiber for forward and backward directions, respectively.~$\varepsilon_{a(b)j}$~and~$\xi_{lj}$~($l$=$L,R$)~are excitation amplitudes of $a_j~(b_j)$~and the $j$th QD with respect to the transition from~$|g_j\rangle$ to~$|e_{lj}\rangle$, respectively. The spatial dependence of wave functions can be expressed as
\begin{align}\label{eqexpmuts}
\Phi_{k,R}(x) = &e^{ikx}[\theta(-x)+a\theta(x)\theta(d-x)\nonumber\\
&+t\theta(x-d)],\\
\Phi_{k,L}(x) = &e^{-ikx}[r\theta(-x)+b\theta(x)\theta (d-x)],
\end{align}
where~$\theta(x)$~is the unit step function, which is equal to 1 when~$x\geq0$~or 0 when~$x<0$.~$e^{ikx}a\theta(x)\theta(d-x)$~and~$e^{-ikx}b\theta(x)\theta(d-x)$~are wave functions of photon between two WGM resonators for forward and backward directions, respectively.~$t$~and~$r$~are transmission and reflection amplitudes, respectively. According to the eigenvalue
equation~$H|\Psi\rangle=E_{k}|\Psi\rangle$, we can analytically obtain a set of coefficients for forward and backward directions, as
\begin{align}\label{e004}
r_f&=\frac{2ih\eta({\rm{B}}e^{2i\theta}{\rm{C^A_-+AC^B_+}})}
{4{\rm{AB}}e^{2i\theta}h^2\eta^2+{\rm{C^A_+C^B_+}}},\\
r_b&=\frac{2ih\eta({\rm{A}}e^{2i\theta}{\rm{C^B_-+BC^A_+}})}
{4{\rm{AB}}e^{2i\theta}h^2\eta^2+{\rm{C^A_+C^B_+}}},\\
t_f&=\frac{\rm{D^A_-D^B_-}}
{4{\rm{AB}}e^{2i\theta}h^2\eta^2+{\rm{C^A_+C^B_+}}},\\
t_b&=\frac{\rm{D^A_+D^B_+}}
{4{\rm{AB}}e^{2i\theta}h^2\eta^2+{\rm{C^A_+C^B_+}}},
\end{align}
where~${\rm{A(B)}}=(\Delta+i\gamma)^2-\omega^2_{1(2)},~{\rm{C^{A(B)}_{\pm}}}=-g^4+{\rm{A(B)}}[h^2+(\gamma\pm\eta)^2]+2i g^2(\Delta+i\gamma)(\gamma\pm\eta),
~{\rm{D^{A(B)}_{\pm}}}=g^4-{\rm{A(B)}}(h^2+\gamma^2-\eta^2)-2ig^2(\Delta\gamma+i\gamma^2\pm\eta\omega_{1(2)})$, detuning $\Delta=\omega-\omega_0$ and $\eta_=G^2/v_g$. $\theta=kd$~denotes phase shift between two WGM resonators. The forward (backward) transmission and reflection are represented by $T_{f(b)}=|t_{f(b)}|^2$  and $R_{f(b)}=|r_{f(b)}|^2$, respectively.
\begin{figure}[t]
\includegraphics[width=3.4in]{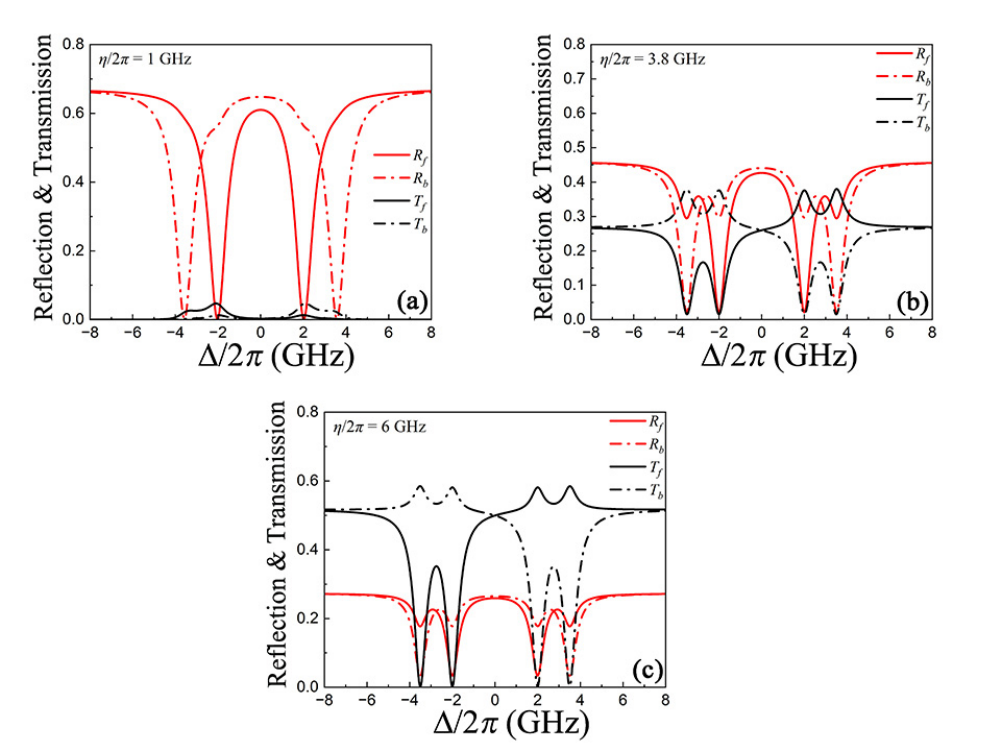}
\caption{\label{2}~Reflections (red lines) and transmissions (black lines) as the functions of detuning $\Delta$ for forward (solid lines) and backward (dot-dashed lines) directions when (a) $\eta/2\pi=1$ GHz, (b) $\eta/2\pi=3.8$ GHz and (c) $\eta/2\pi=6$ GHz. The other parameters are set as $\omega_1/2\pi=2$ GHz, $\omega_2/2\pi=3.5$ GHz, $\theta=\pi$, $g/2\pi=h/2\pi=1$ GHz and $\gamma/2\pi=0.2$ GHz.}
\end{figure}
\section{Results and discussion}
\begin{figure}[t]
\includegraphics[width=3.2in]{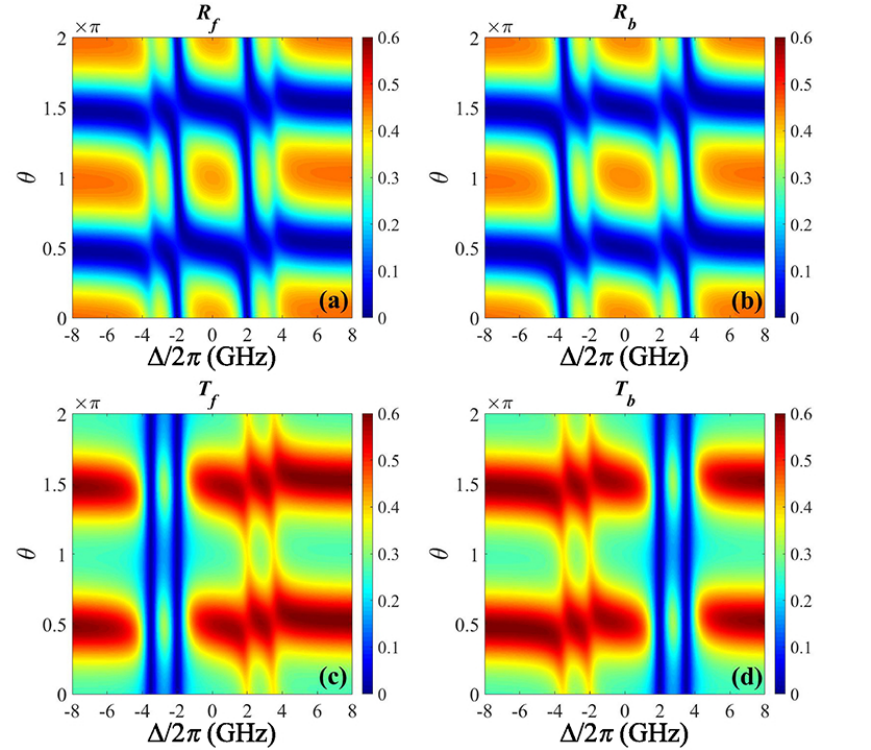}
\caption{\label{3}~Reflections (a), (b) and transmissions (c), (d) versus detuning $\Delta$ and phase shift $\theta$ between two WGM resonators for forward and backward directions, respectively. The other parameters are set as $\omega_1/2\pi=2$ GHz, $\omega_2/2\pi=3.5$ GHz, $\eta/2\pi=3.8$ GHz, $g/2\pi=h/2\pi=1$ GHz and $\gamma/2\pi=0.2$ GHz .}
\end{figure}

First, we show the reflections and transmissions of the system versus detuning $\Delta$ for different coupling strengths~$\eta$~in Fig.\ref{2}.
It can be observed from Fig.\ref{2} (a) that reflections are~$\sim0$ in vicinity of $\Delta/2\pi=\pm2$ GHz for forward direction and $\pm3.5$ GHz for backward direction, respectively, when $\eta/2\pi=1$ GHz.
Obviously, dual-band UR is realized, whereas transmissions in both directions are tiny compared to reflections. This is to say, the system is dominated by UR in this case. As shown in Fig.\ref{2} (b), the forward (backward) transmissions are $\sim0$ in vicinity of~$\Delta/2\pi=-2$~($2$)~GHz and $-3.5$~($3.5$)~GHz, respectively, when $\eta/2\pi=3.8$ GHz. It indicates that the system can also achieve dual-band UT. Moreover, the contrast of reflections for forward and backward directions
~($|R|=|R_f-R_b|$)~is relatively large, and so is the contrast of transmission~($|T|=|T_f-T_b|$). Obviously, the system can realize both UR and UT
under this condition.
It is quite different from the system that UR\cite{Zou2021,Wu2018,Qiu2019} can be implemented but has the same transmissions in both directions.
When $\eta/2\pi=6$ GHz, $|R|$ decreases faster and $|T|$ increases more rapidly in Fig.\ref{2} (c) than it in Figs.\ref{2} (a) and \ref{2} (b), indicating that the system is dominated by UT under this condition. A comparation of Figs.\ref{2} (a), \ref{2} (b) and \ref{2} (c) reveals a gradual decrease in $|R|$ and increase in $|T|$ with increasing of the coupling strength between the WGM resonators and fiber.
\begin{figure}[t]
\includegraphics[width=3.2in]{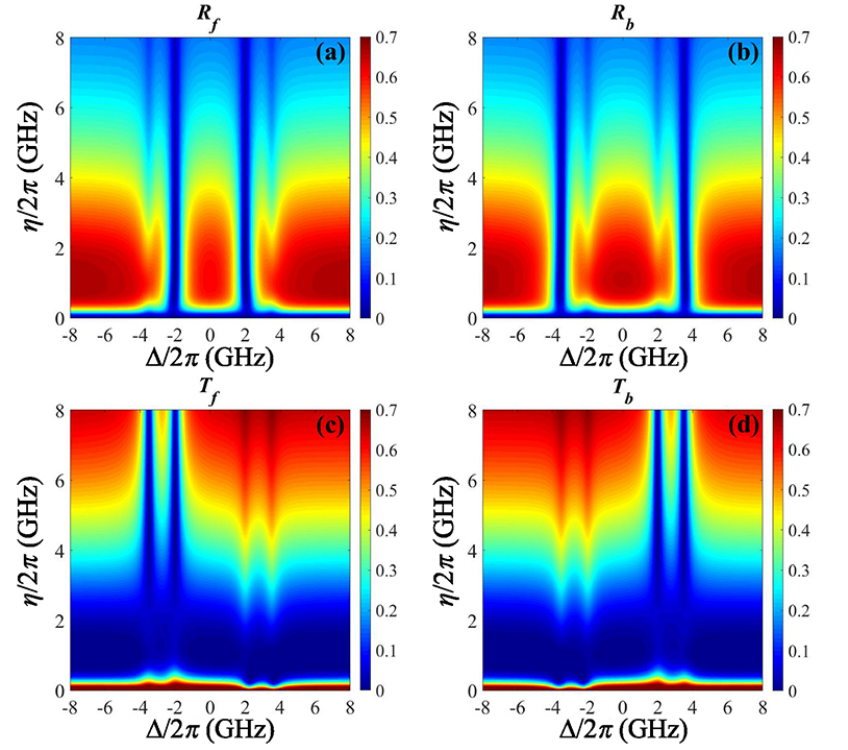}
\caption{\label{4}~Reflections (a), (b) and transmissions (c), (d) as the functions of detuning $\Delta$ and coupling strength $\eta$ between WGM resonators and optical fiber for forward and backward directions, respectively. The other parameters are set as $\omega_1/2\pi=2$ GHz,
$\omega_2/2\pi=3.5$ GHz, $\theta=\pi$, $g/2\pi=h/2\pi=1$ GHz and $\gamma/2\pi=0.2$ GHz.}
\end{figure}

Next we discuss the influences of detuning $\Delta$ and phase shift $\theta$ on reflection and transmission as shown in Fig.\ref{3}. From Figs.\ref{3} (a) and \ref{3} (b), low reflection areas appear around $\Delta/2\pi=\pm2$ and $\pm3.5$ for forward and backward
directions, respectively, when $\theta$ is in range of $0\sim0.2\pi$, $0.8\pi\sim1.2\pi$ and
$1.8\pi\sim2\pi$. Moreover, low reflection areas near $\Delta/2\pi=\pm2~(\pm3.5)$~for forward (backward) direction correspond to high reflection areas for backward (forward) direction. From Fig.\ref{3} (c) (\ref{3} (d)), the low transmission for
forward (backward) direction is shown around $\Delta/2\pi=-2~(2)$ and $-3.5~(3.5)$, respectively. The low transmission areas for
forward (backward) direction correspond to high transmission areas for backward (forward) direction.
Not only that, the low reflection peaks in both directions exhibit weak red shifts within a period accompanying with the increase in phase shift $\theta$, while the low transmission peaks remain almost unchanged with varying in $\theta$, which makes them intersect at $\theta=\pi$ (corresponds to the situation in Fig.\ref{2} (b)). Visibly, UR and
UT can both be realized when the values of detuning and phase shift are appropriate.

\begin{figure}[t]
\includegraphics[width=3.2in]{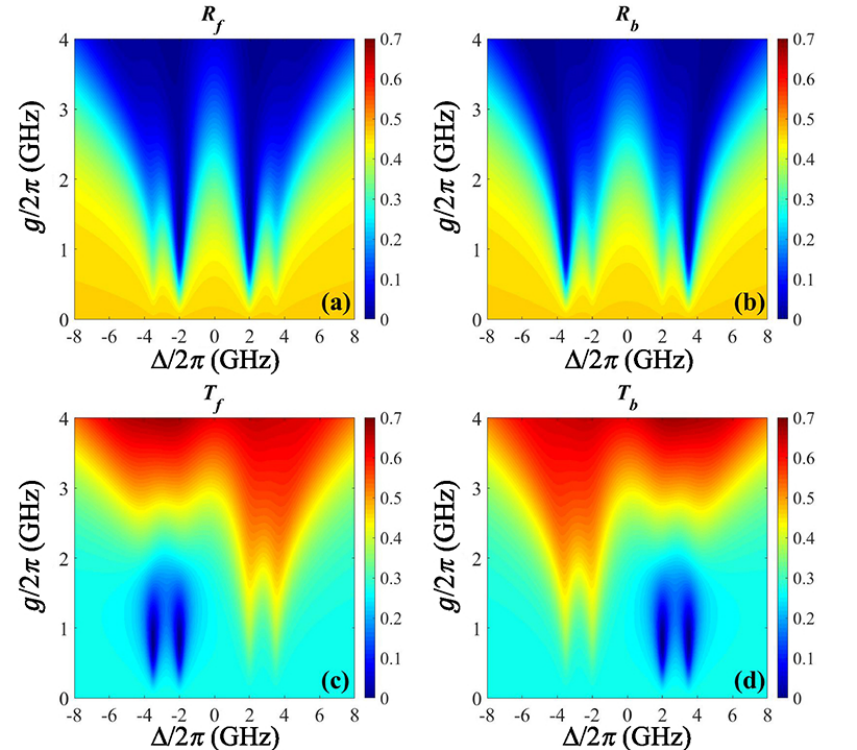}
\caption{\label{5}~Reflections (a), (b) and transmissions (c), (d) versus detuning $\Delta$ and coupling strength $g$
between WGM resonators and QDs for forward and backward directions, respectively. The other parameters are set as $\omega_1/2\pi=2$ GHz, $\omega_2/2\pi=3.5$ GHz, $\eta/2\pi=3.8$ GHz, $\theta=\pi$, $h/2\pi=1$ GHz and $\gamma/2\pi=0.2$ GHz.}
\end{figure}

Then, we discuss the influences of detuning $\Delta$ and coupling strength $\eta$  between WGM resonators and optical fiber on reflection and
transmission. When $\eta/2\pi$ is in range of $0.42~{\rm GHz}\sim2.52$ GHz, as shown in Figs.\ref{4} (a) and \ref{4} (b), it can be found that the low reflection areas for forward (backward) direction
correspond to the high reflection areas for backward (forward) direction around $\pm2~(\pm3.5)$, and the
transmissions for both directions are tiny as shown in Figs.\ref{4} (c) and \ref{4} (d). With increasing $\eta/2\pi$ from
$4.72$ GHz to $7.5$ GHz, the low transmission areas for forward (backward)
direction appear near $-2~(2)$ and $-3.5~(3.5)$, and the low transmission areas for forward (backward)
direction correspond to the high transmission areas for backward (forward) direction. Hence, UR and UT can be realized by adjusting the coupling strength $\eta$ appropriately. In addition, the positions of low reflection peaks and low transmission peaks are immobile with the variation of $\eta$.
\begin{figure}[t]
\includegraphics[width=3.2in]{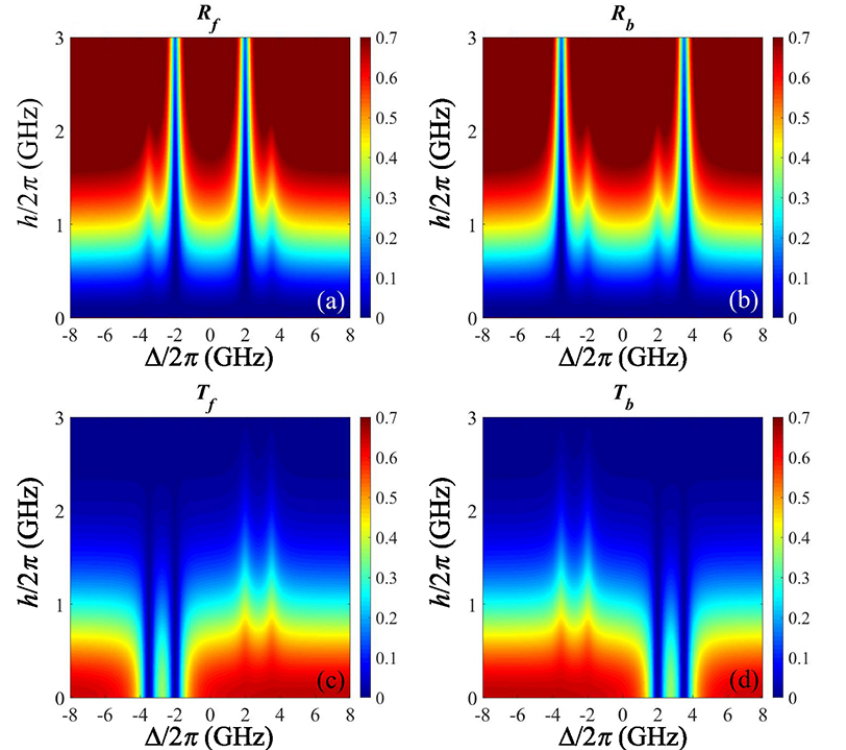}
\caption{\label{6}~Reflections (a), (b) and transmissions (c), (d) versus detuning $\Delta$ and transition rate $h$ between CW mode and CCW mode for forward and backward directions, respectively. The other parameters are set as $\omega_1/2\pi=2$ GHz, $\omega_2/2\pi=3.5$ GHz, $\eta/2\pi=3.8$ GHz, $\theta=\pi$, $g/2\pi=1$ GHz and $\gamma/2\pi=0.2$ GHz.}
\end{figure}

Figure \ref{5} illustrates the effects of  detuning $\Delta$  and coupling strength $g$  between WGM resonators and QDs on reflection and transmission. It can be observed from Figs.\ref{5} (a) and \ref{5} (b) that low reflection areas for forward and backward
directions appear around $\Delta/2\pi=\pm2$ and $\pm3.5$, respectively, when the range of  $g/2\pi$ is approximately $0.8~{\rm GHz}\sim1.2$ GHz. Also in the same range, the low transmission for forward (backward) directions appears around $\Delta/2\pi=-2~(2)$ and $-3.5~(3.5)$,
respectively, according to Figs.\ref{5} (c) and \ref{5} (d). Moreover, the low reflection (transmission) areas for forward direction correspond to the high reflection (transmission) areas for backward direction and vice versa. Therefore, UR and UT can be achieved when $g/2\pi$ is in appropriate range.

Finally, the effects of  detuning $\Delta$ and transition rate $h$  between CW and CCW modes on reflection and transmission are discussed according to Fig.\ref{6}. As shown in
Figs.\ref{6} (a) and \ref{6} (b), the reflections for forward and backward directions are very low when $h/2\pi$ is less than $0.6$ GHz.
With increasing $h/2\pi$ from $0.86$ GHz to $1.4$ GHz, low reflection areas for
forward and backward directions appear around $\Delta/2\pi=\pm2$ and $\pm3.5$, respectively. The low reflection areas for forward (backward) direction correspond to high reflection areas for backward (forward) direction. From Figs.\ref{6} (c) and \ref{6} (d), the forward (backward) low transmission areas around $\Delta/2\pi=-2~(2)$
and $-3.5~(3.5)$ correspond to the backward (forward) high transmission areas, and the obvious UT occurs when $h/2\pi$ is less than $\sim1.4$ GHz. Hence, UR and UT can both be realized when $h/2\pi$ is in range of $0.86$ GHz $\sim1.4$ GHz. In addition, the positions of low
reflection and low transmission peaks are almost unchanged with the varying $h/2\pi$.

\section{Conclusion}
In summary, we have demonstrated the UR and UT properties in a system consisting of two WGM resonators embedded with a Zeeman split QD, respectively, and indirectly coupled through an optical fiber. Research results show that the positions of low reflection peaks exhibit periodic red shifts with the increasing phase shift $\theta$ and intersect with the low transmission peaks simultaneously when phase shift $\theta=\pi$. In other words, both UR and UT can be achieved when $\theta=\pi$. In addition,
UR and UT  can be manipulated by adjusting the coupling strength $\eta$ between WGM resonators and optical fiber as well as the transition rate $h$ between CW mode and CCW mode, while the positions of low reflection and low transmission peaks are robust to the changing of $\eta$ and $h$, but alterable to energy level splitting of QDs. It is worth mentioning that the intermode conversion caused by defective workmanship is fixed during WGM resonator manufacturing, while $\eta$ can be adjusted by altering the distance between WGM resonators and optical fiber. Therefore, we prefer to utilize it as a means of switching between UR and UT. We also investigate the impact of  coupling strength $g$ between QD and WGM resonator on reflection and transmission, and identify the specific range in which UR and UT can occur simultaneously.
In brief, the system has the capability to dynamically regulate photon transport, and it could potentially inspire the advancements in quantum optical devices such as isolators, circulators, routers and so on.

\bibliography{reference.bib}
\bibliographystyle{unsrtnat}

\end{document}